# Chemometrics-aided Surface-enhanced Raman spectrometric detection and quantification of GH and TE hormones in blood

*Annah M. Ondieki[1*], Zephania Birech[1], Kenneth A. Kaduki[1], Peter W. Mwangi[2], Moses Juma[1,3], Boniface M. Chege[2,4]*

[1] Laser Physics and Spectroscopy Research Group, Department of Physics, University of Nairobi, P.O Box 30197-00100, Nairobi, Kenya, [2] Department of Medical Physiology, University of Nairobi, P.O Box 30197-00100, Nairobi, Kenya, [3] UNESCO-UNISA Africa Chair in Nanoscience/Nanotechnology, College of Graduate Studies, University of South Africa (UNISA) South Africa, P.O Box 392 UNISA 0003, South Africa, [4]School of health sciences, Dedan Kimathi University of Technology, Private Bag-1014, Dedan Kimathi, Kenya

Corresponding Authors: [*] moraa94annah@gmail.com (OA);

## Abstract

This work describes the use of Surface-Enhanced Raman Spectroscopy (SERS) technique combined with artificial neural network (ANN) models for detection and quantification of growth hormone (GH) and testosterone (TE) hormones in blood of Sprague Dawley (SD) rats. Here, SERS spectra of blood from SD rats injected with GH only, TE only, both GH and TE hormones and non-injected were recorded and analyzed upon 785 nm laser excitation. The drawn blood samples were first mixed with silver nanoparticles (AgNPs) synthesized via ablation in distilled water, and a drop of the resulting mixture applied onto an aluminum wrapped microscope glass slide, then left to air dry. The SERS spectra of blood from SD rats injected with GH only, TE only, both GH and TE hormones and non-injected displayed identical spectral profiles but with bands exhibiting hormone-dependent intensity variations. These bands were centered around 658, 798, 878, 914, 932, 1064, 1190, 1354, 1410, and 1658 $cm^{-1}$. It was also observed, with the aid of PCA loadings plots, that some Raman band intensities varied with time after rat's injection with GH, TE separately and with both GH and TE. These bands were those centered around wavenumbers 1378 $cm^{-1}$ (for all groups); 658 $cm^{-1}$ and 1614 $cm^{-1}$ (for GH injected rats), 798 $cm^{-1}$ (separately GH and TE injected rats); 786 $cm^{-1}$ (non-injected rats and rats injected with both GH and TE); 914 and 1240 $cm^{-1}$ (TE injected rats and rats injected both GH and TE); 876 and 1636 $cm^{-1}$ (rats injected both GH and TE). The band intensity variations suggested subtle but distinct biochemical changes induced by hormone injections. The, GH and TE ANN models (with six PCA scores of blood spiked with different hormone concentrations as inputs) trained and validated were noted to possess high coefficients of determination ($R^2$) (greater than 87.71%) and low root mean square error (RMSE) values (less than 0.6436). The results showed that hormone-injected rats' respective hormone levels elevated for some time and declined later when compared to the non-injected rats. The same result was observed when ELISA kits were used. Although the two techniques provided similar results, SERS offers additional benefits of being rapid (about two minutes), using simple sample preparation, small sample amounts, and not hormone-specific. This implies that it is possible to detect exogenous injection of sport dopants using SERS spectral data combined with ANN model. These results widen the potential use of SERS in sports science, clinical diagnostics, and biomedical research.

## 1. Introduction

Surface Enhanced Raman spectroscopy (SERS) is a Raman variant technique well-suited for analyzing the sample's molecular composition and structural changes quantitatively (Otange *et al.*, 2017). It also allows the evaluation of multiple indicator parameters in a single measurement, requiring minimal sample preparation (Chege *et al.*, 2019; Birech *et al.*, 2017). The technique allows for the high sensitivity and specificity identification of tiny molecular changes by providing distinct fingerprint information about sample molecules (Ondieki *et al.*, 2022). Although SERS holds significant potential for label-free detection of bio-molecular changes, interpreting the complex spectral information from biological samples presents numerous challenges (Bell *et al.*, 2020; Wang and Yu, 2015). For instance, its quantification procedure is cumbersome as SERS intensity is influenced by not only concentration but also instrumental effects such as spatial resolution and detector sensitivity (Bell *et al.*, 2020). Thus, quantifying chemicals, molecules, or substances using SERS spectral data is often difficult unless chemometric techniques like principal component analysis (PCA) and artificial neural networks (ANNs) are employed (Jones *et al.*, 2019; Zhu *et al.*, 2018; Weng *et al.*, 2015).

SERS combined with chemometrics has thus shown potential in biomedical fields in which hormone analysis such as estrogen (Ondieki *et al.*, 2022; Liu *et al.*, 2018), disease diagnosis such as cancer (Alvarez-Puebla and Blanco-Formoso, 2020; Galata *et al.*, 2019), and drug discovery (Wang *et al.*, 2021) has been achieved. In addition, since this technique provides quick and nondestructive measurements for biological sample examination (Birech *et al.*, 2017), it is very promising for a variety of in-lab and on-site applications such as sport doping detection (Jafari *et al.*, 2021). In detection of sport dopants, this technique overcomes most of the demerits faced by conventional techniques such as enzyme-linked immunoassay (ELISA) (Ahmetov *et al.*, 2019; Ribeiro de Oliveira Longo Schweizer *et al.*, 2018) and chromatographic methods (Carey, 2018). Therefore, there is a great need to develop novel methods for singular or multiple detection of these dopants. In this work, SERS technique combined with ANN has been utilized for the detection of GH and TE hormones in blood from Sprague Dawley (SD) male rats.

## 2. Materials and Methods

Twenty-four SD rats, fed and kept as described in (Ondieki *et al.*, 2023), were divided into four groups: Non-injected rats, GH-injected rats, TE-injected rats, and rats injected with both GH and TE, with each group containing six rats. The rats received intramuscular injections of GH and TE either individually or simultaneously once a day, based on their weight. Blood samples were collected from each rat's orbital sinus at six time points: before injection, and at 30 minutes, 2 hours, 4 hours, 8 hours, and 24 hours after administration. This was to help in determining how long these hormones get elevated in blood before they go back to their normal concentrations. 30 µl of each of the blood samples obtained were mixed thoroughly with 150 µl of AgNPs. 2 µl of the resulting mixture was then dropped onto aluminum-wrapped microscope glass slides (25.4 mm × 76.2 mm × 1.2 mm dimensions) and left to dry in air at room temperature for sixty (60) minutes. The SERS (Raman) measurements were performed on each sample using a portable Raman spectrometer (EZRaman-N Portable Analyzer System, Enwave Optronics, USA) as described in (Ondieki *et al.*, 2023).

Two ANN models were trained and validated based on the PC scores of simulate samples (blood with different concentrations of GH and TE ranging from 0.01 ng/ml to 60 ng/ml) obtained from the PCA process as described in (Ondieki *et al.*, 2023). The input layer of these two models was PC scores and the output was predicted concentration. The hidden layer was made of 5 layers (arranged as 12:10:10:10:6 (for GH model) and 10: 8: 8: 8:6 (for TE model)) with rectifier (ReLU) activation function and resilient backpropagation (rprop+) algorithm. The samples used were about 30 different concentrations (each having 30 spectra) with 75% forming the training set and 25% making the test set. The test for the accuracy of these models was done using the validation metrics such as Root Mean Squared Error (RMSE), and coefficient of determination ($R^2$) determined as shown in equations (i) and (ii) respectively.

$$RMSE = \sqrt{(\frac{1}{N}\sum_{j=1}^{N}(P_j - A_j)^2)} \tag{i}$$

$$R^2 = 1 - \frac{\sum_{j=1}^{n}(A_j - P_j)^2}{\sum_{j=1}^{n}(A_j - \bar{A}_j)^2} \tag{ii}$$

in which $P_j$ represents the value of the predicted concentration, $A_j$ the known concentration value, N the total number of samples, $\bar{A}_j$ represents the average value of the known concentration. The detection limits, specifically the limit of detection (LOD) and limit of quantification (LOQ), were also evaluated. The LOD was determined using equations (iii) (for minimum LOD) and (iv) (for maximum LOD) (Allegrini and Olivieri, 2016).

$$LOD_{min} = 3.3 * [(\sigma_b)^2(S_0)^{-2} + h_{0min}(\sigma_b)^2(S_0)^{-2} + h_{0min}(\sigma_{ycal})^2] \tag{iii}$$

$$LOD_{max} = 3.3 * [(\sigma_b)^2(S_0)^{-2} + h_{0max}(\sigma_b)^2 * (S_0)^{-2} + h_{0max} * (\sigma_{ycal})^2] \tag{iv}$$

in which $\sigma_b$ represents the standard deviation of the response (SERS spectra of male rat's blood), $S_0$ represents the sensitivity of the calibration sample with minimum analyte concentration, $\sigma_{ycal}$ is the standard deviation of calibration errors (residues), and $h_{0min}$ and $h_{0max}$ are the minimum and maximum blank leverages. The value 3.3 is the expansion factors obtained assuming a 95% confidence interval. The minimum and maximum LOQ were calculated using the two respective equations (Eqn. iii and iv) by replacing the value 3.3 with 10 (Desimoni and Brunetti, 2015; Steyn *et al.*, 2011). Low values of minimum detection limits suggested good model performance. On ensuring that the models were accurate, the models were then used to predict the concentrations of these hormones (GH and TE) in blood from rats not injected and those injected with GH only, TE only, and both GH and TE. All these was done in MATLAB R2021a environment (version 9.10.0.1602886, The MathWorks Inc., Natick, USA).

## 3. Results and Discussions

### *3.1. Qualitative analysis of blood from injected and non-injected rats*

To identify the Raman intensity bands associated with GH, TE and both hormones (GH and TE) in blood of SD rats, the averaged SERS spectra of blood from SD rats injected with GH only, TE only, both hormones and non-injected rats is displayed in Figure 1.

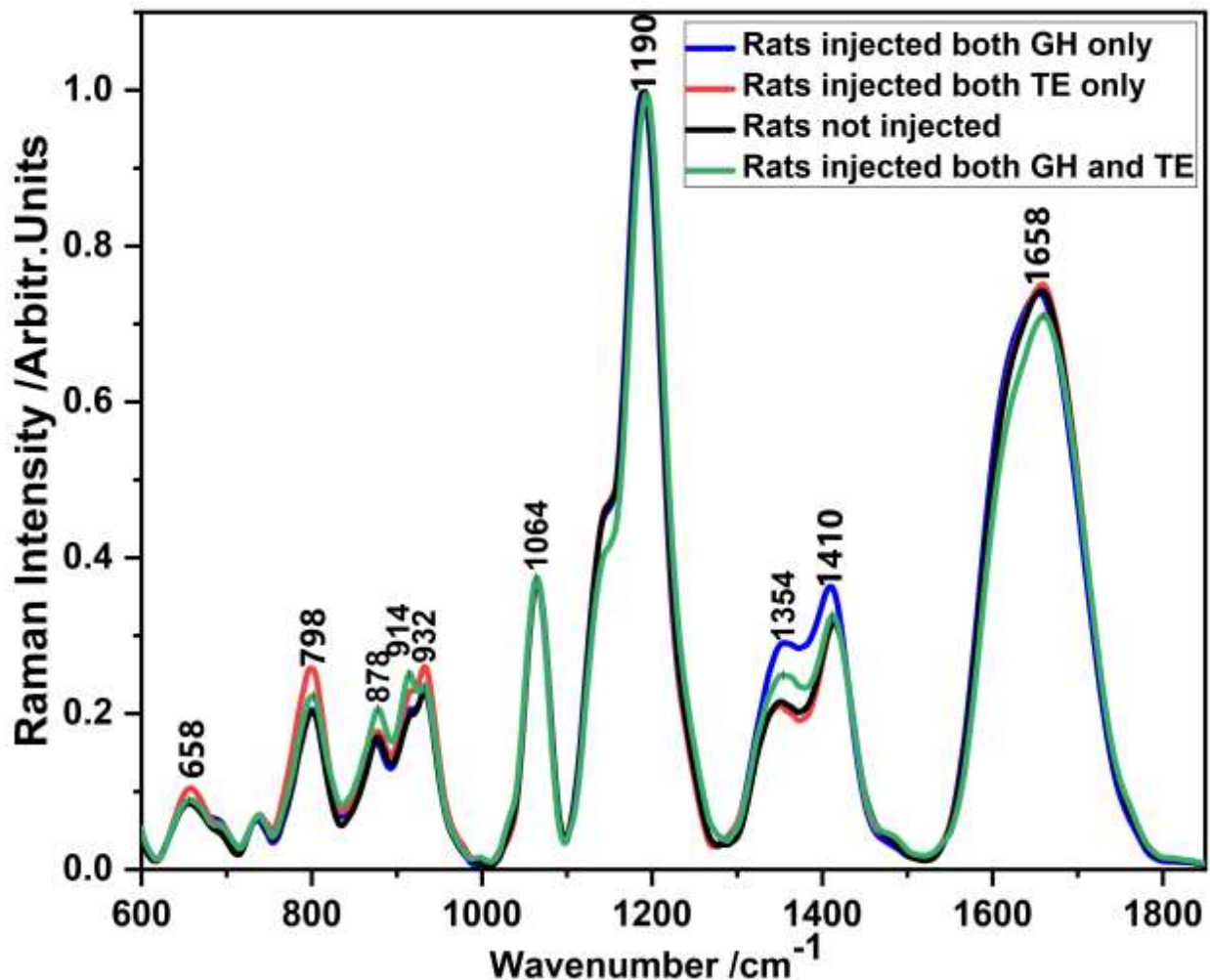

*Figure 1: SERS spectra for blood from rats injected with GH only, TE only, both GH and TE hormones, and non-injected rats*

The SERS spectra of blood from non-injected rats and the injected, as shown in Figure 1, are nearly identical, with a few minor band intensity variations suggesting subtle but distinct biochemical changes induced by hormone injections (Pataky *et al.*, 2021). The prominent bands noted were those centered around wavenumbers 658, 798, 878, 914, 932, 1064, 1190, 1354, 1410, and 1658 cm$^{-1}$. These bands were found to relate to those obtained from blood, and blood mixed with various concentrations of these hormones (Ondieki *et al.*, 2023). The bands centered around 600 to 1064 cm$^{-1}$ were assigned to C-C stretching (Ondieki *et al.*, 2022; Zhou *et al.*, 2020; Barkur and Chidangil, 2019; Jalaja *et al.*, 2017), 1190 cm$^{-1}$ assigned to C-O Stretch and COH bending (Bonnier and Byrne, 2012), 1354 cm$^{-1}$ assigned to Stretch (COO-) and C-H bend (Zhou *et al.*, 2020), and 1658 cm$^{-1}$ assigned to C-O/C=O Stretching (Ondieki *et al.*, 2022; Otange *et al.*, 2017). The SERS bands with the highest intensity were those centered at around 658, 798, 932, and 1658 cm$^{-1}$ for blood from rats injected with TE only; 878 and 914 cm$^{-1}$ for blood from rats injected with both hormones; and 1354 and 1410 cm$^{-1}$ for blood from rats injected with GH only. This analysis underscores the sensitivity of SERS in detecting hormonal influences on blood composition, providing a valuable tool for studying the biochemical effects of hormone treatments.

PCA was used to further investigate if there were differences in spectral patterns of the blood from non-injected rats and those injected with each of the two hormones. Figure 2 shows the PCA scores and loadings plots for blood from GH-injected rats, TE-injected rats, and non-injected rats.

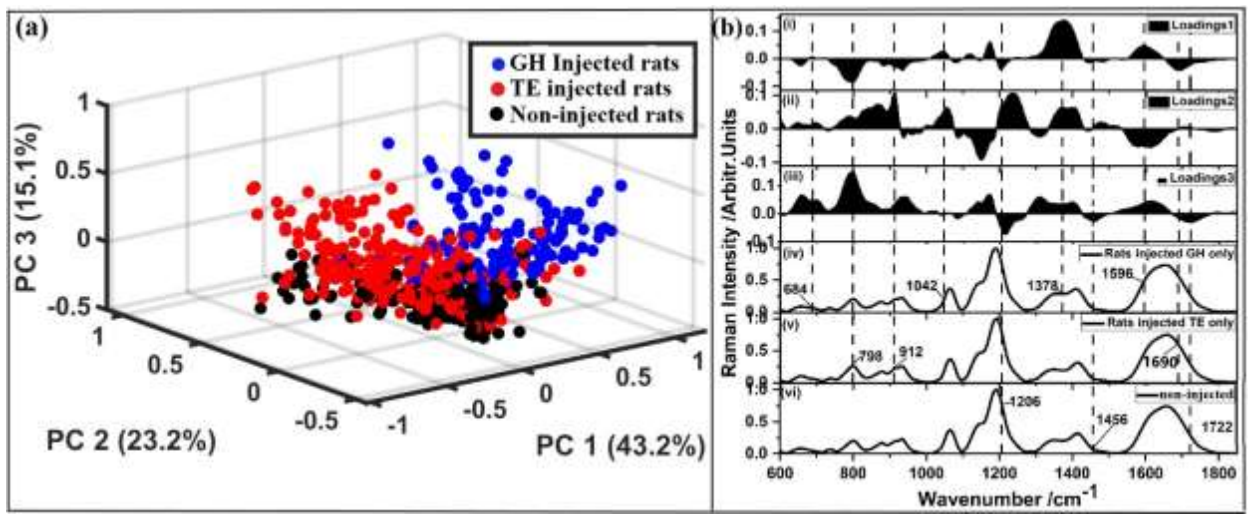

*Figure 2: Three-dimensional PCA (a) score plots, (b) loadings plot for (i) PC1, (ii) PC2, and (iii) PC3 loadings plot, and averaged SERS spectra of blood from (iv) rats injected with GH only, (v) rats injected with TE only and (vi) non-injected rats. The explained variances are indicated in percentages and were 43.2%, 23.2%, and 15.1% for PC1, PC2, and PC3 respectively.*

The displayed score plots show clear segregation of the SERS spectral data sets for blood from GH-injected rats, TE-injected rats, and non-injected rats thus supporting the idea that the SERS spectroscopic technique is a potentially sensitive alternative hormonal assaying method (see Figure 2a). The three PCs explained 81.5% of the variability of the data, where PC1, PC2, and PC3 accounted for 43.2%, 23.2%, and 15.1% respectively. The SERS bands that brought about this segregation were determined using PCA loadings plot based on the PCA score segregation (see Figure 2.b.(i), 2.b.(ii) and 2.b.(iii)). These bands are those centered at wavenumbers 684, 1042, 1378, and 1596 cm$^{-1}$ for GH-injected rats (see Figure 2.b.(iv)); 798, 912, and 1690 cm$^{-1}$ for TE-injected rats (see Figure 2.b.(v)); and 1206, 1456 and 1722 cm$^{-1}$ for non-injected rats (see Figure 2.b.(vi)). Some of these bands matched exactly with some bands identified as concentration sensitive bands using simulate samples as shown by (Ondieki *et al.*, 2023).

To check whether there was spectral variation for different times before and after hormone injection, the average SERS spectra for each of the groups was plotted for different times - before injection and 0.5, 2, 4, 8, and 24 hours after injection as shown in Figure 3. From Figure 3, it is seen that the spectral profiles are very similar in each of the hormones for different times before and after injection. This likely means that some molecular constituents or metabolic pathways have relatively stable spectral signatures against the background of dynamic metabolic processes in response to hormone injection. Although the SERS spectral profiles obtained were generally very similar, subtle variations in SERS band intensities at specific wavenumbers of different times of the day and hormone groups are apparent. Noticeably, bands centered at 1350 and 1410 cm$^{-1}$ show apparent differences related to non-injected rats, GH-injected rats, and TE-injected rats. Also, the band at 798 cm$^{-1}$ has some variation, specifically with the GH injection. This may be an indication of the fact that these variations in metabolic responses of the hormone-specific molecular rearrangements within the tissue microenvironment mean an elevation of these hormones after an injection.

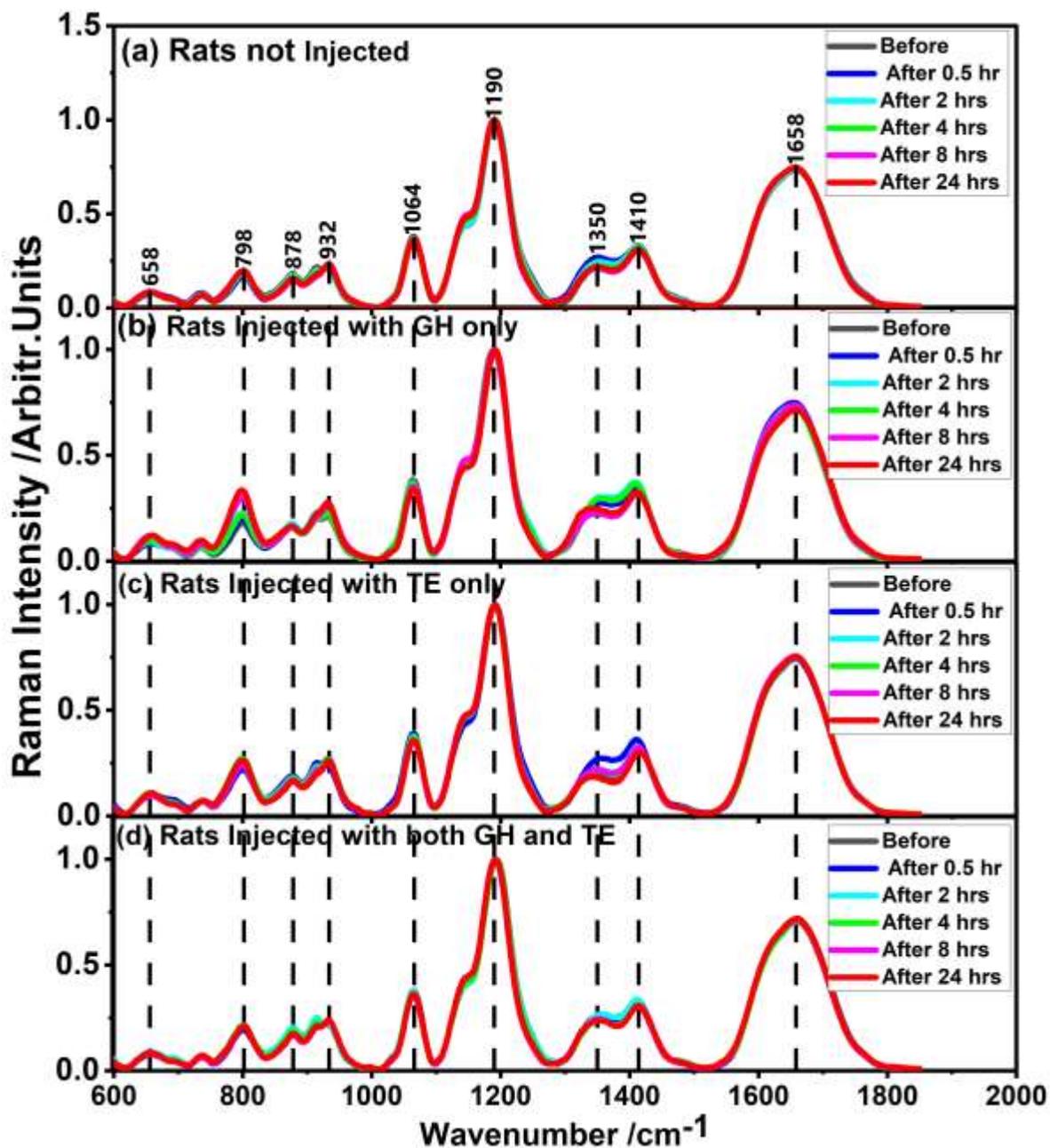

*Figure 3: SERS spectra of different for blood from (a) non-injected rats, (b) rats injected with GH only, (c) rats injected with TE only, and (d) rats injected with both GH and TE hormones. The blood was drawn at different times before and after injection.*

To determine further if there was spectral variation for different times before and after hormone injection, PCA was performed for each group for each at different times (before injection and 0.5, 2, 4, 8, and 24 hours after injection) (see Figure 4).

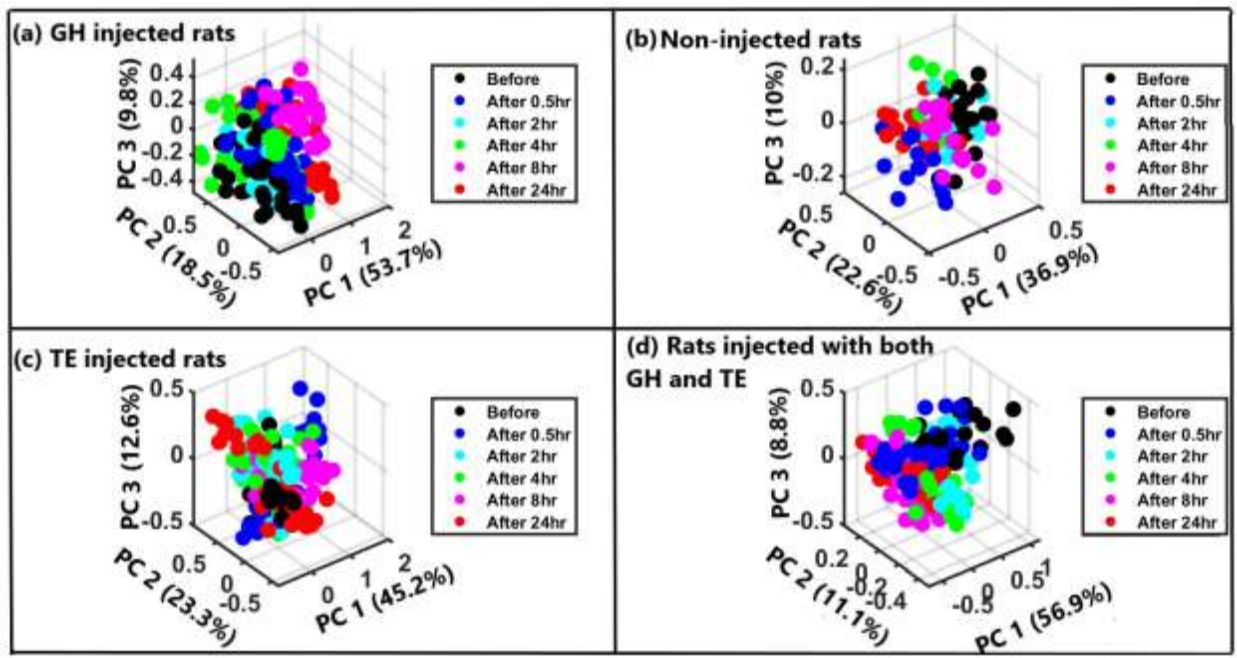

*Figure 4: Three-dimensional PCA score plots for (a) rats injected with GH only, (b) Normal rats, (c) rats injected with TE only, and (d) rats injected simultaneously with GH and TE based on different times before and after injection.*

As displayed in Figure 4, the first three PCs explained 82% (GH injected rats), 81.1% (TE injected rats), 76.8% (GH + TE injected rats), and 69.5% (non-injected rats) of the variance suggesting that the spectral profiles of blood obtained at different times were different. GH-injected rats had the highest variance, closely followed by TE-injected rats, while variance reduced in rats injected simultaneously with both hormones when compared to those injected singly. Non-injected rats exhibited the least variance, indicating more consistent spectral profiles.

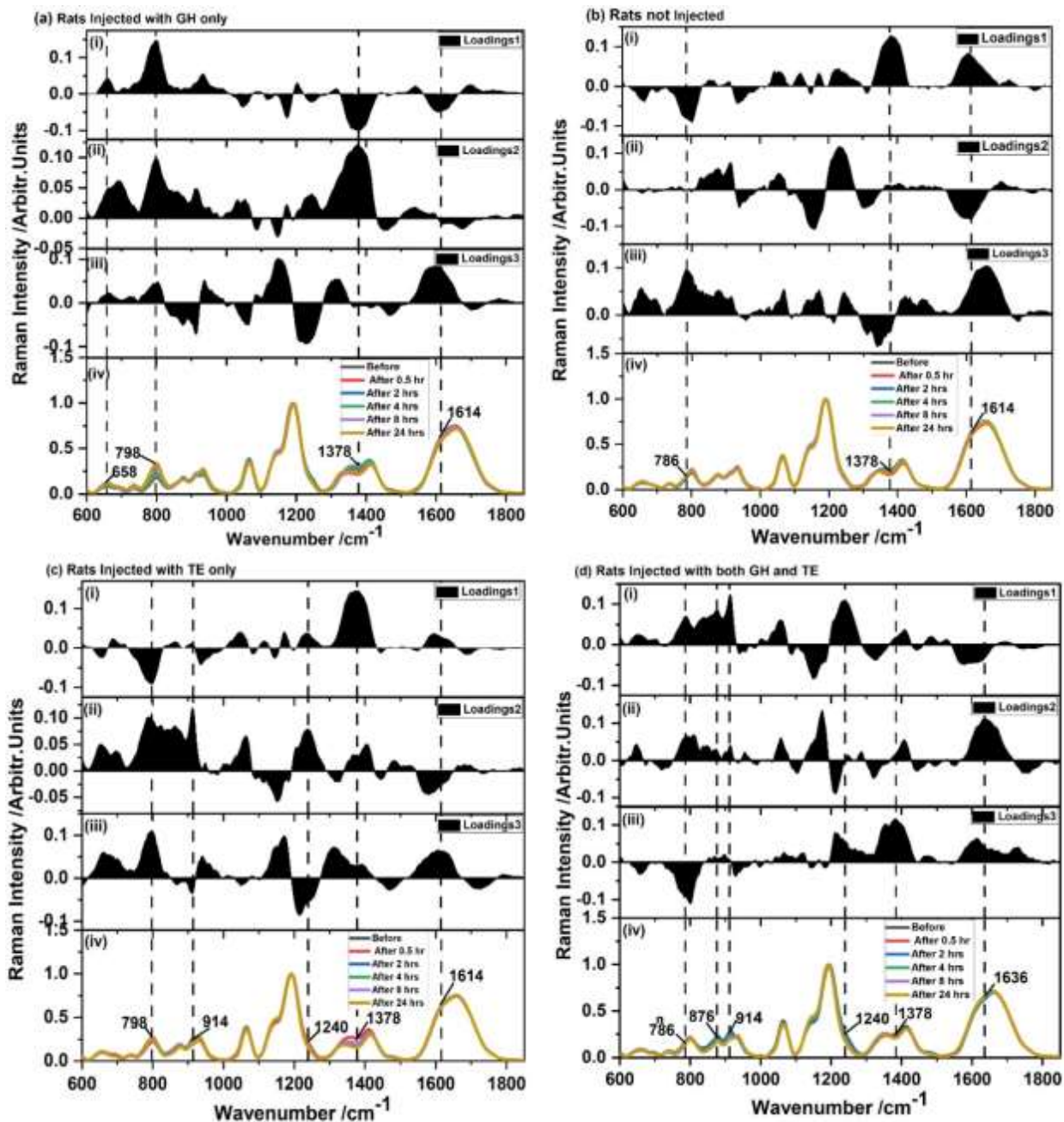

*Figure 5: PCA loading plots and SERS Spectra of blood from rats injected with (a) GH only, (b) non-injected rats, (c) rats injected with TE only, and (d) rats injected simultaneously with GH and TE hormones drawn at different times before and after injection.*

The loadings plot displayed in Fig. 5 shows that the bands responsible for the segregation of the data sets in the score plots were centered at around 1378 cm$^{-1}$ (for all groups); 658 cm-1, 1614 cm$^{-1}$ (for GH injected rats), 798 cm$^{-1}$ (separately GH and TE injected rats); 786 cm$^{-1}$ (non-injected rats and rats injected with both GH and TE); 914 and 1240 cm$^{-1}$ (TE injected rats and rats injected both GH and TE); 876 and 1636 cm-1 (rats injected both GH and TE). These bands could be used as Raman biomarker bands for periodic concentration (level) changes in blood for the respective hormones. These results also point to the great power of the SERS method in detecting subtle respective hormone level changes in blood.

*3.2. Quantitative analysis of GH and TE hormones*

Blood is a complex matrix composed of a myriad number of Raman active biological molecules with isoenergetic vibrational bands. In order to quantify the levels of the hormones of interest in this work, GH and TE, in blood multivariate analytical methods are necessary. In this work, artificial neural network (ANN) were employed in performing concentration determination in the investigated blood samples. Two ANN models (one for TE and the other for GH) were trained, validated and used to predict each hormone's concentration in blood. Table 1 displays validation results obtained from these developed ANN models.

*Table 1: Validation metrics of the ANN models constructed, trained and validated.*

| Model | RMSE (ng/ml) | | $R^2$ | |
|---|---|---|---|---|
| | Training | Validation | Training | Validation |
| Growth Hormone | 0.54489 | 0.64364 | 0.9135 | 0.8771 |
| Testosterone | 0.23870 | 0.42386 | 0.9804 | 0.9331 |

The success of the model is normally assessed by the high (low) values of the coefficient of determination, $R^2$ (low values of the root-mean square error, RMSE). Indeed, as seen from Table 1, the RMSE values were low and the $R^2$ values were close to one which was an indication that these models were suitable for performing level predictions of these hormones in blood using the SERS data sets. Further performance evaluation of the models were assessed using the regression plots of predicted concentration versus the known (see Figure 6).

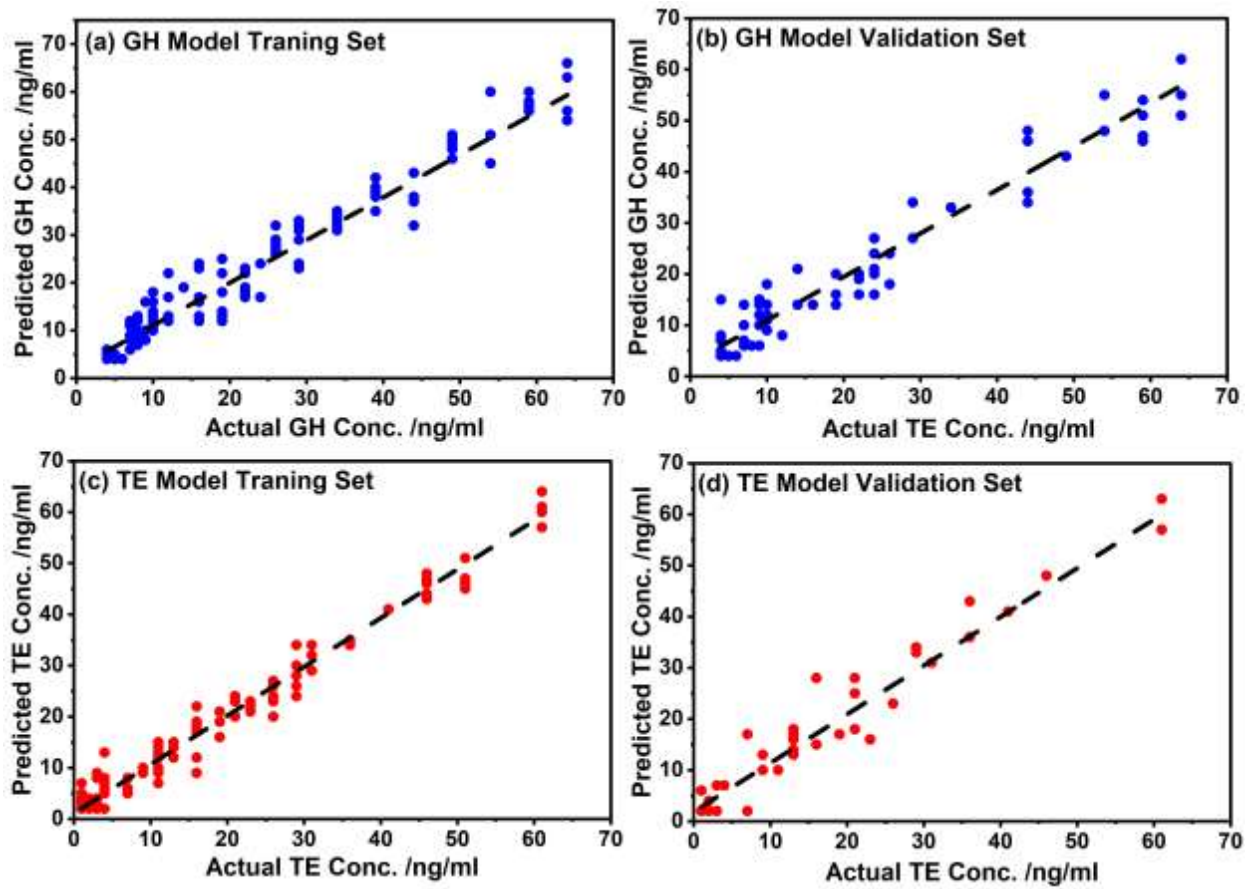

*Figure 6: ANN Models' regression curves showing predicted concentration against actual concentration.*

In all models, data were close to the regression lines revealing that the predicted concentration of both the training and testing set was in acceptable good agreement with the experimental (actual) concentration. This confirmed the excellent performance of the models.

Using equations iii and iv evaluated were also the detection limits i.e., LOD and LOQ of the SERS method used in our work (see Table 2).

*Table 2: The comparison between the detection limits for the ANN models created in this work with those published in the literature.*

| Hormone | Limit of Detection (LOD) (ng/ml) | | Limit of Quantification (LOQ) (ng/ml) | | Reported LOD in literature (ng/ml) | Reported LOQ in literature (ng/ml) | References |
|---|---|---|---|---|---|---|---|
| | $LOD_{min}$ | $LOD_{max}$ | $LOQ_{min}$ | $LOQ_{max}$ | | | |
| Growth hormone | 0.3222` | 42.33 | 0.9766 | 128.28 | 0.04 (ELISA) and 0.5 (IDMS) | 0.065 (ELISA) and 0.7 (IDMS) | (Miller *et al.*, 2022; Steyn *et al.*, 2011; Arsene *et al.*, 2010) |
| Testosterone | 0.1851 | 18.57 | 0.5611 | 56.28 | 0.2 (LC-HRMS) and 0.05 - 0.2 (NanoLC-HRMS/MS) | 0.5 (LC-HRMS) | (Jing *et al.*, 2022; Solheim *et al.*, 2022) |

The minimum LOD deduced were 0.3222 ng/ml (for GH) and 0.1851 ng/ml (for TE) while the minimum LOQ determined were 0.9766 ng/ml (for GH) and 0.5611 ng/ml (for TE). This revealed that the LOD and LOQ values obtained for TE enanthate were notably in range with those reported by other analytical methods, as evidenced by the comparative data presented in Table 2. For instance, from previous studies done on detection of TE using dried blood samples, the LOD obtained ranged from 0.05 to 0.2 ng/ml when using NanoLC-HRMS/MS technique depending on the type of TE injected intramuscularly (Solheim *et al.*, 2022). Similarly, a LOD of 0.2 ng/ml and a LOQ of 0.5 ng/ml for TE enanthate was obtained when using dried blood samples extraction coupled with TurboFlow technology and LC-HRMS by means of the DBSA-TLXHRMS system (Jing *et al.*, 2022). This implies that the SERS method combined with ANN models offers superior sensitivity in detecting and quantifying TE enanthate, thereby enhancing the precision and reliability of measurements in related research and clinical applications. When using GH ELISA assays the LOD and LOQ obtained were 0.04 and 0.065 ng/ml respectively (Miller *et al.*, 2022; Steyn *et al.*, 2011), while using isotope dilution mass spectrometry (IDMS) the values were 0.5 ng/ml (LOD) and 0.7 ng/ml (LOQ) (Arsene *et al.*, 2010). Although the GH minimum detection limits obtained in this study are seen to be higher when compared to ELISA and lower when compared to IDMS (Table 5.3) these values (LOD and LOQ) may still be acceptable since they are very low when compared with the range of GH concentrations typically encountered in the rat blood samples (3-25 ng/ml). Upon ascertaining that these models were accurate and had acceptable detection limits, they were used to detect the levels GH and TE hormones in blood of rats injected singly and simultaneously with the two hormones in comparison to those not injected. This is because elevation in concentrations of these hormones in blood above the normal or known concentrations might suggest doping which is often done by athletes to increase their sport activity (Holt and Ho, 2019). The average values of the predicted concentration levels of the respective hormones were determined before and some hours after injection (0.5, 2, 4, 8, and 24 hours) in each of the groups (injected and non-injected). Figure 7 displays GH and TE hormone levels (on average) in blood taken from GH with/and TE injected and non-injected SD rats throughout the 24 hours of study.

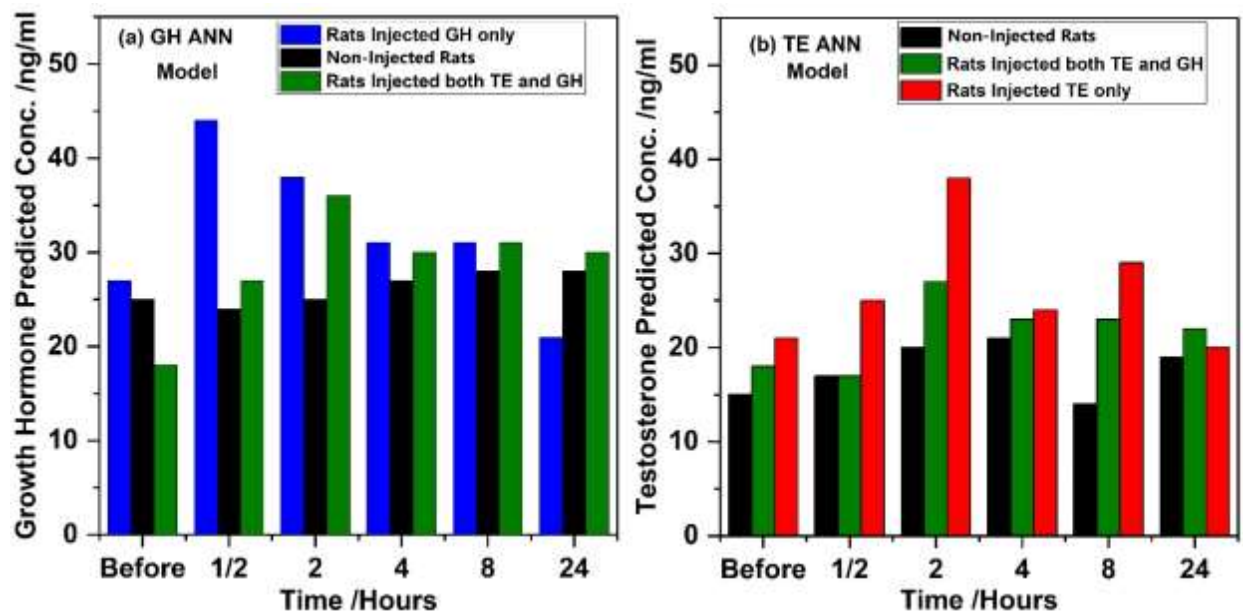

*Figure 7: Predicted concentrations using the (a) GH and (b) TE ANN models*

The administration of GH led to a rapid elevation in GH levels within half an hour, reaching a peak concentration of 44 ng/ml, indicating the prompt bioavailability and uptake of exogenous

GH. This elevation in GH levels was sustained for a brief period before gradually declining, although remaining elevated compared to baseline levels (Figure 6(a)). The return of GH levels to normal after 24 hours suggests the transient nature of exogenous GH effects, with hormonal homeostasis being restored over time. Similarly, administration of TE resulted in a progressive increase in TE levels over time, peaking between half an hour and eight hours post-injection (Figure 6(b)), consistent with the expected pharmacokinetic profile of TE. Interestingly, concurrent administration of TE and GH did not significantly alter the temporal pattern of TE elevation compared to TE alone, indicating independent mechanisms of action for these hormones (Gibney *et al.*, 2005). Importantly, TE levels returned to baseline after 24 hours across all treatment groups, underscoring the transient nature of TE effects and the efficient metabolism and elimination of exogenous TE. The observed changes in hormone levels highlight the dynamic interplay between exogenously administered hormones and endogenous hormonal regulation mechanisms. The rapid elevation and subsequent decline in GH levels reflect the pulsatile nature of GH secretion and the regulatory feedback mechanisms that maintain hormonal balance (Holt and Ho, 2019). Similarly, the temporal profile of TE levels reflects its pharmacokinetic properties, including absorption, distribution, metabolism, and elimination processes.

To validate the data obtained, ELISA was used and the results are displayed in Figure 8. Here ELISA standard curves of each hormone was plotted as a graph of optical density against the known concentration as in figure 8(a) for GH and 8(b) for TE. A best-fit curve, i.e., a Boltzmann function (for GH) and exponential decay function (for TE), was then applied to these points to establish a relationship between concentration and optical density. This fit allowed for prediction of the concentrations of GH and TE hormones in unknown blood samples by comparing their optical density to the standard curve (see Figure 8(c) and 8(d)).

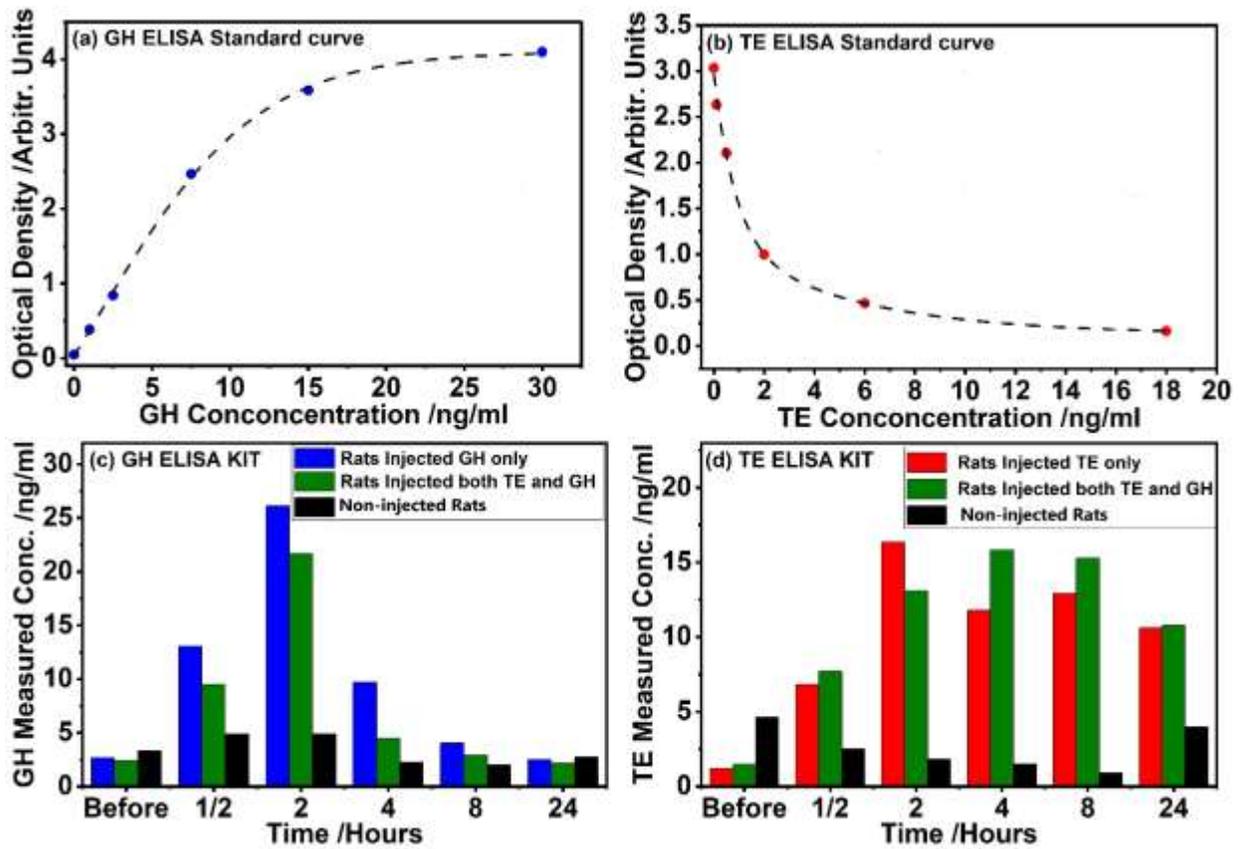

*Figure 8: ELISA results for non-injected rats and rats injected singly and simultaneously with GH and TE hormones.*

As seen from the standard curves, the data showed an excellent fit, with the coefficient of determination values being nearly one i.e., 0.99946 for GH (Figure 8(a)) and 0.99733 for TE (Figure 8(b)). The predicted concentration by ELISA, was noted to correlate with those concentrations obtained by the ANN models. Initially, GH concentrations in both non-injected rats and rats injected with hormones were within normal ranges (3 ng/ml), indicating baseline physiological levels. However, following GH injection, GH concentrations rapidly increased, reaching peak levels (26 ng/ml) within two hours post-injection, before returning to baseline levels by four hours. This temporal profile suggests a transient elevation in GH concentrations, with exogenously administered GH being detectable in the bloodstream for up to two hours post-injection. Beyond this timeframe, GH concentrations reverted to normal levels, underscoring the limited duration of exogenous GH detection using the ELISA kit employed in our study. Conversely, TE levels at baseline were also within normal ranges in both non-injected rats and rats injected with hormones (2 ng/ml). Following TE injection, TE concentrations gradually increased, peaking at two hours for rats injected with TE alone and at eight hours for rats injected simultaneously with GH and TE. This sustained elevation in TE concentrations suggests a prolonged presence of exogenously administered TE in the bloodstream, with detectable levels lasting up to eight hours post-injection. Importantly, these findings are consistent with results obtained from SERS with ANN models, providing additional validation of the ELISA kit results.

The accuracy and dependability of various analytical methods for measuring hormone concentrations after exogenous injection can be better understood by comparing the results of ELISA and SERS with ANN. In this study, significant similarities as well as differences between the two approaches' temporal profiles of GH and TE concentrations were noted. As demonstrated by the findings of the ELISA and SERS with ANN, the injection of GH triggered a rapid and sustained rise in GH levels within the first two hours after the injection. The GH concentration that peaked between 30 minutes and 2 hours after injection and then gradually decreased is consistent with the pharmacokinetic profile of GH that is given exogenously. This consistency between SERS with ANN and ELISA results underscores the reliability of both techniques in capturing the dynamic changes in GH concentrations over time. Similarly, the administration of TE led to a progressive increase in TE levels over time, peaking between half an hour and eight hours post-injection, as observed with both SERS with ANN and ELISA methods. The sustained elevation in TE concentrations and the prolonged detection window of up to eight hours post-injection are consistent across both analytical techniques, reflecting the pharmacokinetic properties of exogenously administered TE. Additionally, both SERS with ANN and ELISA results showed that simultaneous injection of TE and GH did not significantly change the temporal patterns of TE concentrations compared to TE alone. It indicates that TE and GH have separate modes of action, with neither hormone significantly interfering with the other's ability to affect hormone concentrations. The validity of SERS with ANN in capturing the temporal changes of GH and TE concentrations is therefore demonstrated by the consistency of the hormone quantification results between SERS with ANN and ELISA technique. This demonstrates that the combination of SERS with ANN provides a label-free, non-destructive approach for multiplexed hormone measurement, with possible advantages in terms of sample throughput and adaptability.

## 4. Conclusion

This work has demonstrated the ability of SERS together with ANN models in quantifying the GH and TE hormone levels in blood of male SD rats. Here, two ANN models were developed based on six PC scores obtained from the PCA of different concentrations of each of the hormones in blood. A test for the model's accuracy was done using $R^2$ and RMSE values that were seen to be greater than 87.71% and less than 0.6436 respectively implying the models were accurate. The minimum limit of detection (LOD) deduced were 0.3222 ng/ml (for GH) and 0.1851 ng/ml (for TE) while the minimum limit of quantification (LOQ) determined were 0.9766 ng/ml (for GH) and 0.5611 ng/ml (for TE). This detection limits were very low and thus acceptable. Using the calibrated ANN models in determining the concentration hormone level in blood of rats, it was noted that hormone injected rats' respective hormone levels elevated for some time and declined later when compared to those of non-injected rats. This implied that exogenous injection of sport dopants can be detected and quantified using SERS technique combined with ANN models. This could bring about the customization of a SERS system that utilizes SERS data and inbuilt ANN models to detect GH and TE levels in blood in less than a minute. These findings widen the potential use of SERS in sports science, clinical diagnostics, and biomedical research.

**Acknowledgments**

We sincerely express our gratitude to the Swedish International Development Cooperation Agency (SIDA) through the International Science Programme (ISP), Uppsala University, for sponsoring this research.